\documentclass[12pt]{article}
\pdfoutput=1
\usepackage[top=1in, bottom=1in, left=1.in, right=1.in]{geometry}
\usepackage[english]{babel}
\usepackage{atbegshi,cite}
\usepackage{amsmath,amssymb,amsbsy,amstext, amsthm, simplewick}
\usepackage{hyperref}
\usepackage{graphicx}
\usepackage{amsfonts}
\usepackage[small]{caption}
\usepackage{upgreek}
\usepackage[titletoc]{appendix}
\usepackage{setspace}
\usepackage{color}

\newcommand{\newc}{\newcommand}
\newc{\fpi}{f_{\pi}}
\newc{\etap}{\eta^{\prime}}
\newc{\llll}{\langle\lambda\lambda\rangle}
\newc{\FFd}{F^a\tilde F^a}
\newc{\qbar}{{\overline q}}
\newc{\TR}{{\rm Tr}}
\newc{\Kahler}{K\"ahler }
\newc{\Zbb}{{\mathbb Z}}
\newc{\Rt}{{\mathbb R}^3}
\newc{\Rf}{{\mathbb R}^4}
\newc{\So}{{\mathbb S}^1}
\newc{\zt}{{\mathbb Z}_2}
\newc{\RtSo}{{\mathbb R}^3\times{\mathbb S}^1}
\newc{\scriminus}{{\cal I}^-}
\newc{\scriplus}{{\cal I}^+}
\newc{\mpl}{M_p}
\newc{\Ricci}{\mathcal{R}}
\newc{\bv}{\phi}
\newc{\calU}{{\cal U}}
\newc{\calK}{K}
\newc{\calUi}{{\cal U}^{-1}}
\newc{\calG}{{\cal G}}
\newc{\calO}{{\cal O}}
\newc{\calQ}{{\cal Q}}
\newc{\calOb}{{\cal O}^\dagger}
\newc{\hphi}{{\hat\phi}}

\theoremstyle{plain}
\theoremstyle{plain} 
\theoremstyle{plain} 
\theoremstyle{plain}
\theoremstyle{plain}
\theoremstyle{plain}

\renewcommand{\title}[1]{{\Large\bf\flushleft{#1}}\vspace*{3ex}\\}
\renewcommand{\author}[2]{{\noindent\hspace*{2.5em}\large#1}
                     \footnote{Electronic mail: $\mathtt{#2}$}\\}

\begin{document}
\begin{titlepage}
\begin{flushright}
{\large 
~\\
}
\end{flushright}

\vskip 2.2cm

\begin{center}

{\large \bf Comments on the CKN Bound}

\vskip 1.4cm

{{Tom Banks}$^{(a)}$ and {Patrick Draper}$^{(b)}$}
\\
\vskip 1cm
{$^{(a)}$ Department of Physics and NHETC, Rutgers University, Piscataway, NJ 08854}\\
{$^{(b)}$ Department of Physics, University of Illinois, Urbana, IL 61801}
\vspace{0.3cm}
\vskip 4pt

\vskip 1.5cm

\begin{abstract}
Cohen, Kaplan, and Nelson (CKN) conjectured that the UV and IR cutoffs of effective quantum field theories coupled to gravity are not independent, but are connected by the physics of black holes. We interpret the CKN bound as a scale-dependent depletion of the QFT density of states and  discuss various aspects of the bound on small  and large scales. For laboratory experiments, we argue that the bound provides small corrections to ordinary quantum field theory, which we estimate to be of order $m_e/M_p$ for $g-2$ of the electron. On large scales, we suggest a modification of the CKN bound due to the presence of cosmological horizons and discuss the connection with entropy bounds. 
\end{abstract}

\end{center}

\vskip 1.0 cm

\end{titlepage}
\setcounter{footnote}{0} 
\setcounter{page}{1}
\setcounter{section}{0} \setcounter{subsection}{0}
\setcounter{subsubsection}{0}
\setcounter{figure}{0}



\section{Introduction}
Cohen, Kaplan, and Nelson~\cite{ckn} pointed out that UV and IR cutoffs on local effective quantum field theory might be correlated in the presence of gravity.
Placing an ordinary EFT in a box of size $L$, CKN suggested that the UV cutoff $\Lambda_{UV}$ should be low enough that states of characteristic energy density $\Lambda_{UV}^4$ are not black holes:
\begin{align}
r_b(\Lambda_{UV}, L) < L~~~\Rightarrow ~~~\Lambda_{UV}^4L^3 < L M_p^2
\end{align}
or
\begin{align}
\Lambda_{UV} < \sqrt{M_p/L}
\label{eq:CKN}
\end{align}
up to order one numbers.

It  is  obvious that  EFT becomes suspect whenever gravitational backreaction is large enough to hide the box behind an event horizon. However, merely excluding ``black hole states" from field theory has negligible impact on ordinary perturbative processes with small numbers of particles.\footnote{Apart from rendering the perturbative series somewhat better defined. States of energy $M$ in regions of size $R$ with entropy $S>MR$ are problematic because $e^Se^{-MR}>1$.  The Bekenstein bound excludes these states, but the CKN bound is even stronger, amounting to $S_{EFT}<R^{3/2}$.} The CKN bound is much stronger than this. 
It can also be stated as a conjecture about the EFT density of states. Ordinarily, in a box of volume $V=L^3$,  the number of states in a momentum space volume $d^3p$ is 
\begin{align}
dN\sim Vd^3p. 
\end{align}
This leads, for example, to the ``quartic divergence in the vacuum energy density,"
\begin{align}
\Delta E/V \sim V^{-1} \int d\epsilon\, \epsilon D(\epsilon) \;,\;\;\;\; D(\epsilon)\sim V\epsilon^2
\end{align}
summing over zero-point energies of the single-particle states.
The CKN bound can be interpreted as a conjecture that the density of states that is well-described by quantum field theory is severely depleted for energies $\epsilon>\sqrt{M_p/L}$,  behaving as though the box is of size $L(\epsilon) < M_p/\epsilon^2$, or
\begin{align}
dN\sim M_p^3\epsilon^{-6}d^3p 
\label{eq:DOS}
\end{align}
saturating the inequality.

Taking $L=H^{-1}$ in an expanding universe of Hubble parameter $H$, the ``bare vacuum energy density" is is of order $H^2M_p^2$. Modes of energy lower than $\sqrt{HM_p}$ appear to give a quantum correction to the vacuum energy density which is of order the  bare value. Modes of energy higher than $\sqrt{HM_p}$ contribute to the usual fine-tuning problem. It is these modes that are depleted by the CKN bound, and it was suggested in~\cite{ckn}  that the low UV cutoff on ordinary QFT might explain why the cosmological constant is not quartically sensitive to the highest energy scales through radiative corrections. With the QFT density of states~(\ref{eq:DOS}) high energy modes contribute
\begin{align}
\Delta E/V &\sim H^{3} \int_{\sqrt{HM_p}}^\infty d\epsilon\, M_p^3 \epsilon^{-3}\nonumber\\
&\sim H^2M_p^2.
\label{eq:vacenergy}
\end{align}
Although this result does not explain the large size of the universe, it might resolve the fine-tuning problem. Holographic arguments support a similar conclusion~\cite{Thomas:2002pq}.

\section{Modification on Cosmological Scales}
\label{sec:cos}
The CKN bound as stated above requires some modification on cosmological scales. (In this section we set $M_p=1$.) Exciting finite-energy states over large scales changes the cosmological horizon. For example, in de Sitter space with dS length $\ell$, 
there is a cosmological horizon at $r_c=\ell$ in static coordinates. Therefore one might like to consider EFT in boxes as large as $L=r_c$ and place a CKN-type bound on the UV cutoff. However, there are no black holes in this asymptotic spacetime with radius larger than the Nariai size $r_N=\ell/\sqrt{3}$. 

To resolve this puzzle, we can extend the CKN bound to
\begin{align}
r_b(\Lambda_{UV}, L;\ell) < L <  r_c(\Lambda_{UV}, L;\ell).
\label{eq:rbLrc}
\end{align}
Here $r_b$ and $r_c$ are the black hole and cosmological horizons of the Schwarzschild-de Sitter spacetime with dS length $\ell$ and mass $m\equiv\Lambda_{UV}^4L^3$. 
The SdS horizons are two of the three solutions to $1-\frac{2m}{r}-\frac{r^2}{\ell^2}=0$. Therefore, both conditions~(\ref{eq:rbLrc}) can be imposed at once  by requiring
\begin{align}
1-\frac{2m}{L}-\frac{L^2}{\ell^2}>0,
\end{align}
which leads to
\begin{align}
\Lambda_{UV} < \left(\frac{1}{2L^2}-\frac{1}{2\ell^2}\right)^{\frac{1}{4}}.
\label{eq:CKNmod}
\end{align}
The bound~(\ref{eq:CKNmod}) reduces to that of CKN for $L\ll r_N$. On larger scales, it encodes the requirement that  EFT should not include states for which the cosmological horizon is inside the box. In particular, for $L\rightarrow\ell$, the UV cutoff goes to zero. 

We can give two interpretations of this result. We might treat the bound~(\ref{eq:CKNmod}) as an effective modification of the density of states, as above. In this case we again obtain a zero-point contribution of order the bare scale, $\Delta E/V\sim 1/\ell^2$. 
On the other hand, the vanishing UV cutoff may instead imply that there is {\em{no}} renormalization of the cc from conventional EFT.  
This is consistent with substantial evidence accumulated by Banks and Fischler that the cc should be thought of as a fixed parameter~\cite{Banks:2018jqo}. A similar conclusion was recently drawn by Bramante and Gould~\cite{Bramante:2019uub}.  
We will mainly suggest an interpretation of~(\ref{eq:CKNmod}) in relation to entropy bounds.

\section{Relation to Entropy Bounds}
It was argued in~\cite{Banks:2019oiz} that the CKN bound arises from a form of the covariant entropy principle~\cite{Fischler:1998st,ceb} applied to field-theoretic degrees of freedom,
\begin{align}
S_{\diamond, EFT} <c\, A_{\diamond}^{\frac{3}{4}} ,
\label{eq:ceb}
\end{align}
where $\diamond$ denotes a causal diamond, $S_{\diamond, EFT}$ is the entropy associated with degrees of freedom that can be well-approximated by local EFT inside the diamond, and $A_{\diamond}$ is the area of the holographic screen, the maximal-area 2-surface on the boundary of the diamond. $c$ is a number of order one.
For small causal diamonds, the entropy in local QFT degrees of freedom is of order $(L\Lambda_{UV})^3$, so  replacing $A_{\diamond}\rightarrow L^2$, we obtain
\begin{align}
(L\Lambda_{UV})^3 \lesssim (L^2)^{\frac{3}{4}} ,
\end{align}
recovering the CKN bound~(\ref{eq:CKN}).

\begin{figure}[t!]
\begin{center}
\includegraphics[width=0.5\linewidth]{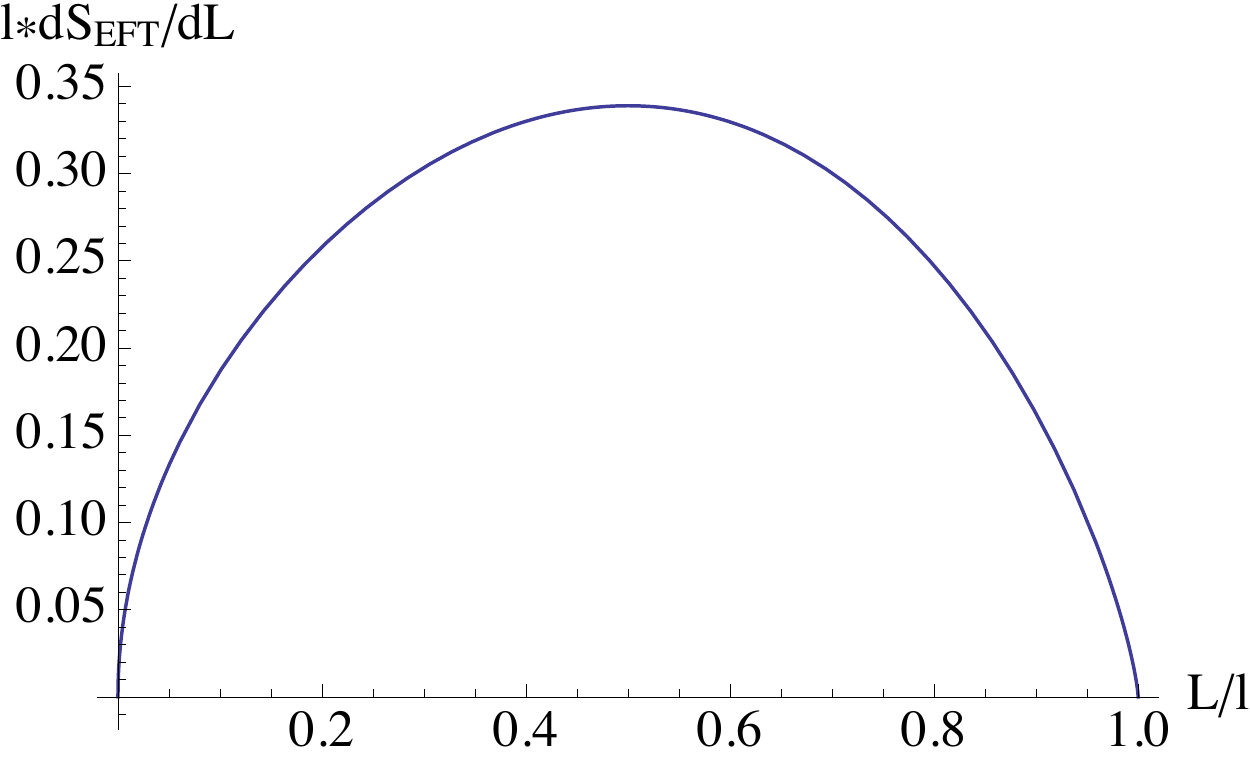}
\caption{Most of the states that can be well-described by QFT are localized well inside the cosmological horizon.
} 
\label{fig:dSdL}
\end{center}
\end{figure} 
More generally, in the static patch, the entropy in states that can be well-described by local QFT degrees of freedom (at least over times of order $\ell$) increases with the box size as
\begin{align}
dS_{EFT} \simeq (\Lambda_{UV}(L))^3L^2 dL\;.
\label{eq:dSEFT}
\end{align}
If the UV cutoff on states of characteristic size $L$ does not fall too quickly with $L$, we recover $S_{EFT}\simeq (L\Lambda_{UV}(L))^3$. This is the case for the ordinary CKN bound~(\ref{eq:CKN}). However, if $\Lambda_{UV}(L)$ is taken to saturate~(\ref{eq:CKNmod}), the estimate $S_{EFT}\simeq (L\Lambda_{UV}(L))^3$ is invalid for $L$ of cosmological size. Indeed, it predicts $S_{EFT}\rightarrow 0$. Instead, (\ref{eq:dSEFT}) indicates that $dS_{EFT}/dL$ is peaked at $L=\ell/2$ and goes to zero as $L\rightarrow \ell$. $dS_{EFT}/dL$ is shown in Fig.~\ref{fig:dSdL}. 

In other words, in the largest causal diamonds, most of the entropy that can be well-described by  local QFT degrees of freedom over times of order $\ell$ corresponds to states that are localized an ${\cal O}(1)$ distance inside the diamond, rather than roughly homogeneous and horizon-sized. This is what we learn about field theory entropy from~(\ref{eq:CKNmod}) and is consistent with the arguments of~\cite{Banks:2010tj}.

Integrating Eq.~(\ref{eq:dSEFT}) and replacing $L^2\rightarrow A_\diamond$, we find
\begin{align}
S_{\diamond, EFT} <f(L/\ell)\, A_{\diamond}^{\frac{3}{4}} ,
\end{align}
where $f$ is a function that always of order one for $L$ between zero and $\ell$. We find that the modified CKN bound~(\ref{eq:CKNmod}) still qualitatively saturates the $S<A^\frac{3}{4}$ bound on field theory entropy.

\section{g-2}
CKN argue that their bound is conceivably testable in precision measurements, and they estimate the effects of correlated UV and IR cutoffs on the one-loop contribution to $g-2$ of the electron,
\begin{align}
\delta(g-2)\sim\frac{\alpha}{\pi}\left[\left(\frac{m_e}{\Lambda_{UV}}\right)^2+\left(\frac{1}{m_e L(\Lambda_{UV})}\right)^2\right]
\label{eq:gminus2ckn}
\end{align}
with $L$ and $\Lambda$ related by Eq.~(\ref{eq:CKN}). (See also the recent work~\cite{bramante2019anomalous}.) The estimate~(\ref{eq:gminus2ckn}) comes from placing hard momentum cutoffs on the usual Feynman integral. However, if the correct implementation of the bound is in terms of a scale-dependent density of states, then Eq.~(\ref{eq:gminus2ckn}) is an overestimate. 

The usual one-loop correction to $g-2$ is schematically
\begin{align}
(g-2)\sim e^2 m_e^2 \int_0^1 dxdydz\,  &\delta(x+y+z-1) \int \frac{d\ell^0}{2\pi} \int \frac{d^3\ell}{(2\pi)^3} \frac{z(1-z)}{(\ell^2+\Delta^2)^3},\nonumber\\
&\Delta\equiv(1-z)^2 m_e^2
\label{eq:gm2}
\end{align}
(in Euclidean space, at leading order in the momentum of the external photon, and omitting order-one numbers). 
The integral is dominated by loop momenta of order $m_e$, so a more appropriate estimate of the effects of the scale-dependent density of states is to place the system in a box of size $L(m_e)=M_p/m_e^2$. (This scale is enormous, and so the rest of the exercise is academic, showing only that such corrections are negligibly small.) Finite volume effects can then be estimated with standard techniques (see, e.g.,~\cite{Kronfeld:2002pi}). We use an exponential representation of the propagators and discretize spatial momenta,
\begin{align}
\frac{1}{(\ell^2+\Delta)^3}&\rightarrow \int d\rho\, \rho^2 e^{-\rho( \ell_0^2+\Delta)}e^{-\rho\ell_i^2}\nonumber\\
\ell_i&\rightarrow \frac{2\pi\nu_i}{L}\nonumber\\
\int d^3 \ell_i &\rightarrow \frac{1}{L^3}\sum_{\nu_i}.
\end{align}
Each of the spatial momentum sums can be reorganized with Poisson summation,
\begin{align}
\sum_\nu e^{-\rho(2\pi\nu/L)^2} \rightarrow \frac{L}{\sqrt{\rho}}\sum_n e^{-\frac{1}{\rho}\left(\frac{L n}{4\pi}\right)^2}
\end{align}
again dropping order one numbers. For large $L$ the sum is dominated by $n=0,\pm 1$, where $n=0$  corresponds to the infinite volume limit. We can then perform the integrals over $\ell_0$  and the Feynman parameters  to obtain
\begin{align}
\delta(g-2)_{IR}&\equiv (g-2)|_{Lm_e\gg 1}-(g-2)|_{L\rightarrow\infty}\nonumber\\
&\sim \frac{\alpha}{\pi} \int d\rho\, \left(\frac{e^{-\frac{L^2}{16 \pi ^2 \rho }}}{m_e \rho^{3/2}}+{\cal O}\left(\frac{1}{m_e\sqrt{\rho}}\right)\right)\;.
\end{align}
Since the integral is dominated by $\rho\sim L^2$,  higher-order terms in $1/(\sqrt{\rho} m_e)$ can be dropped. Performing the $\rho$ integral we obtain the leading correction to $g-2$ from a finite box of size $L(m_e)$,
\begin{align}
\delta(g-2)_{IR}\sim \frac{1}{m_eL(m_e)}\sim \frac{m_e}{M_p}.
\end{align}

Another way  to arrive at this result is to note that the modification to $g-2$ is  the difference between doing an ordinary integral in a momentum range around $m_e$ and approximating the integral with a Riemann sum with momentum bins of size $2\pi/L$. The ``error" in the Riemann sum approximation to this region of the integral is linear in $1/L$. 

 Including also an ordinary UV cutoff (without CKN effects) for comparison, we obtain
\begin{align}
\delta(g-2)\sim \frac{\alpha}{\pi}\left[\left(\frac{m_e}{\Lambda_{UV}}\right)^2+\left(\frac{m_e}{M_p}\right)\right].
\label{eq:deltagm2}
\end{align}
We see that the depleted density of states gives results comparable to ordinary UV contributions for $\Lambda\gtrsim \sqrt{m_e M_p}\sim 10^7$ GeV.

We can obtain a similar estimate from the effect of the density of states on corrections to $g-2$ from new physics at $\Lambda_{UV}$  by putting this part of the  momentum integral in  a box of size $L(\Lambda_{UV})$,
\begin{align}
\delta(g-2)_{UV}\sim&\frac{\alpha}{\pi}\left[\left(\frac{m_e}{\Lambda_{UV}}\right)^2\left(1+\frac{1}{\Lambda_{UV} L(\Lambda_{UV})}\right)\right]\nonumber\\
\sim &\frac{\alpha}{\pi}\left[\left(\frac{m_e}{\Lambda_{UV}}\right)^2+\frac{m_e^2}{M_p\Lambda_{UV}}\right].
\end{align}
The first term is the typical contribution from a standard UV cutoff, ignoring CKN corrections, and the second is the typical  CKN correction to this contribution as estimated above.
We see that the second term is of higher order compared to the second term in Eq.~(\ref{eq:deltagm2}); most of the modifications from  a  CKN-type bound  indeed arise from momentum  scales around $m_e$.

\section{Discussion}

We have discussed an interpretation of the CKN bound as a depletion of the QFT density of states, and argued that the usual Schwarzschild black hole argument requires modification on cosmological scales. This modification supports the idea that most field theory states in dS space are localized well inside the cosmological horizon. We  have also estimated the effect of the CKN bound on small scale physics and found that it is negligibly small.

Let us conclude with three additional comments.

\begin{itemize}

\item The impact of the CKN bound on scalar fields is of interest. The density of states~(\ref{eq:DOS}) implies that the 
effective potential for scalar fields computed in ordinary quantum field theory (e.g., the field dependent terms in the vacuum energy~(\ref{eq:vacenergy})) should not be thought of as providing a correction to the cosmological constant.

On the other hand, as our analysis  of $g-2$ indicates, the predictions of  quantum  field theory  for local measurements at SM energies are not substantially modified by the CKN bound. 
In light of this, the effective potential computed in ordinary QFT without the density of states~(\ref{eq:DOS}) should accurately capture loop corrections to on-shell scalar processes with loop momenta $k$ and external momenta  $k_{ext}$ satisfying 
\begin{align}
L(k)^{-1} \ll k_{ext} \ll k. 
\end{align}
The first inequality is a CKN bound and the second from the interpretation of the effective potential as the leading term in a derivative expansion. Thus the usual quadratic divergence in scalar masses is still a problem.  The extent of the electroweak hierarchy problem that can be inferred from ordinary QFT is characterized by  the UV scale 
\begin{align}
\Lambda\sim\sqrt{m_W M_p}, 
\end{align}
at which $L$ becomes shorter than the weak scale and the Higgs no longer fits into the CKN box.

We are quite uncertain about the extent to which EFT calculations of an effective potential correspond to any calculation in a true theory of quantum gravity.  String theory teaches us to extract effective Lagrangians from scattering amplitudes. Sen et. al. and Seiberg~\cite{Seiberg:1986ea,Pius:2013sca,Pius:2014iaa} have shown how to extract one-loop on-shell mass renormalization in this manner, while Kaplunovsky has done the same for gauge coupling renormalization~\cite{vadim}. Our remarks above about on-shell scalar mass renormalization are thus on pretty safe grounds. Interpretation of the effective potential  as a calculable contribution to the cc, however, is quite problematic, as  illustrated by the CKN bound and reinforced by a host of other more rigorous evidence~\cite{Banks:2018jqo}.

\item It is known from tests of the equivalence principle that ``vacuum loops gravitate"~\cite{Polchinski:2006gy}. In brief, the idea is that radiative corrections to the binding energies of different nuclei $A$ and $B$ are much larger than the precision with which $m^A_{\tiny inertial}/m^A_{\tiny gravitational}=m^B_{\tiny inertial}/m^B_{\tiny gravitational}$  has been tested. Therefore, at least in the background fields of a nucleus, loops couple to gravity~\cite{Polchinski:2006gy}. This is not inconsistent with the CKN bound, because the relevant length scales in the radiative contributions to nuclear binding energies are ${\cal  O}({\rm fm})$.  CKN effects are expected to be of order GeV/$M_p$.

\item The thermodynamics of SdS black holes  indicates that general localized excitations in dS space are low-entropy constrained states of the dS thermal bath (see~\cite{Banks:2006rx,Banks:2010tj} and references therein; also more recently~\cite{Johnson:2019vqf,Dinsmore:2019elr,Johnson:2019ayc}.) Most of these states are black holes that do not have a field-theoretic description. In any case, the fact that localized states have low entropy indicates that they are unstable over sufficiently long times. All ``field theory states" are localized and the bounds above should be understood to apply on timescales shorter than the time over which they decay back into the dS bath.

\end{itemize}

It would be interesting to study further the implications of these sorts of bounds for other precision tests and scalar field cosmology.

\vskip 1cm
\noindent
{\bf Acknowledgements:}  The work of PD is supported by NSF grant PHY-1719642. The work of TB is partially supported by the U.S. Department of Energy under Grant DE-SC0010008.

\bibliography{ckn_refs}{}

\bibliographystyle{utphys}

\end{document}